\documentclass[12pt]{iopart}

\usepackage{iopams}
\usepackage{graphicx}
\begin{document}

\title{Properties of electron lenses produced by ponderomotive potential with Bessel and Laguerre--Gaussian beams}

\author{Yuuki Uesugi$^{1,2}$, Yuichi Kozawa$^{1}$ and Shunichi Sato$^{1}$}

\address{$^{1}$ Institute of Multidisciplinary Research for Advanced Materials, Tohoku University, Katahira 2-1-1, Aoba-ku, Sendai-shi, Miyagi 980-8577, Japan}
\address{$^{2}$ PRESTO, Japan Science and Technology Agency, Kawaguchi-shi, Saitama 332-0012, Japan}
\ead{uesugi@tohoku.ac.jp}
\vspace{10pt}
\begin{indented}
\item[]January 2022
\end{indented}

\begin{abstract}
The properties of electron round lenses produced by the ponderomotive potential are investigated in geometrical optics.
The potential proportional to the intensity distribution of a focused first-order Bessel or Laguerre--Gaussian beam is exploited to produce an electron round lens and a third-order spherical aberration corrector.
Several formulas for the focal length and spherical aberration coefficients in the thin-lens approximation are derived to set the lens properties and associated light beam parameters.
When the mode field of the optical beam is small, the electron trajectory calculation results show properties similar to those obtained using the formulas.
Alternatively, large higher-order aberrations are introduced because of the annular distribution of the potential.
The second- and higher-order Bessel and Laguerre--Gaussian beams produce no focusing power and no negative third-order spherical aberration; however, they can still be used as circularly symmetric higher-order aberration correctors.
Results show that the ponderomotive potential--based electron lens or phase plate forms a refractive index medium with a shape that is considerably more flexible than that achieved in the case of conventional electrostatic and magnetic electron optics.
The formulas presented herein can serve as guidelines for designing preferred light fields, thus promoting the advancement of a novel technology in electron optics that exploits the electron--light interaction.
\end{abstract}

%
%
%
%
%

\section{Introduction}
The electron microscope was first invented in the 1930s based on the concept of ultramicroscopy using electrons with a wavelength considerably shorter than that of light.
Subsequently, theoretical studies on the properties of electron microscope lenses---or electron lenses---were conducted to organise the fundamentals of electron optics based on electrostatic and magnetic fields.
Electron lenses substantially differ from optical lenses because their refractive index distribution cannot be freely designed in space.
The electrons do not pass through any electrode or magnetic pole; hence, the distribution of the resulting potential must follow the Laplace equation. 
Consequently, a circularly symmetric electron lens---or an electron round lens---always functions as a convex lens with a positive spherical aberration (SA) coefficient \cite{Scherzer1936} (refer to the Appendix for a discussion on achieving the sign of this coefficient).
Since the inception of electron microscopy, the correction of the SA has been a major concern in the resolution improvement.
However, this issue was only overcome in the 1990s \cite{Zach1995,Haider1998,Krivanek1999}, nearly half a century after Scherzer theoretically demonstrated that SA can be corrected using a combination of multipole lenses \cite{Scherzer1947}.

In the currently available finest electron microscopes, aberration correctors eliminate third- and fifth-order SAs as well as chromatic aberrations.
In scanning transmission electron microscopy, atomic resolution imaging can be achieved using a focused electron beam with a probe size of $\sim0.5$ \r{A} \cite{Erni2009,Sawada2015}.
Such a miniscule probe is realised using a corrector and a monochromator.
The probe size is usually limited to $\sim1$ nm without correction for SAs.
Electron microscopy with atomic resolution imaging is a key technology for promoting cutting-edge material science and quantum engineering.
However, it shows disadvantages---such as high installation costs, the complicated operation and adjustment of the electron optics and poor long-term stability---because aberration correctors necessitate the high-precision control of 12 magnets per multipole lens unit.

For more than a decade now, a new technology that employs an intense light field as an electron optical phase element has garnered considerable attention \cite{Aflatooni2001,Batelaan2007,Hebeisen2008,Muller2010,Handali2015,Vidil2015,Matthias2017,Kozak2018,Talebi2019,Konecna2020,Wang2020,Feist2020,Kozak2021}.
This is accomplished by exploiting stimulated Compton scattering in quantum mechanics or the ponderomotive force in classical physics \cite{Fedorov1997}.
In both cases, the accumulated phase shift is produced by the potential proportional to the square of the electromagnetic field provided by a light field and imprints a specific phase profile on the wavefront of an electron beam.

\begin{figure}[tbp]
    \centering
    \includegraphics[width=73mm]{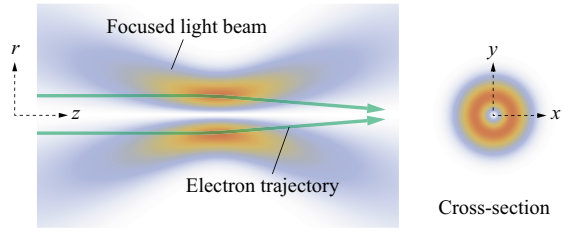}
    \caption{
        Conceptual diagram of a ponderomotive lens (p-lens).
        The first-order Laguerre--Gaussian beam is considered as an example.
        The density plot is the intensity distribution of the light beam.
        An electron beam and the light beam show a common propagation axis.
        Green arrows are electron trajectories (not to scale).
        The right side depicts a cross section of the light beam at the beam waist.
    }
    \label{fig:concept}
\end{figure}

The interaction Hamiltonian between the light and an electron in the Coulomb gauge is expressed as \cite{Axelrod2020,GarciadeAbajo2021}
\begin{eqnarray}
    \fl\mathcal{H}(\mathbf{R},t)=
        e\mathbf{A}(\mathbf{R},t)\cdot\mathbf{v}
        +\frac{e^{2}}{2\gamma m}
            \biggl[
                A_{x}^{2}(\mathbf{R},t)
                +A_{y}^{2}(\mathbf{R},t)
            \biggr.\nonumber\\
            \biggl.
                +\frac{A_{z}^{2}(\mathbf{R},t)}{\gamma^{2}}
            \biggr],\label{eq:interactionH}
\end{eqnarray}
where $\mathbf{A}=(A_{x},A_{y},A_{z})$ denotes the vector potential of the light field and $\mathbf{v},e,m$, and $\gamma=1/\sqrt{1-v^{2}/c^{2}}$ represent the velocity, charge, mass, and Lorentz factor of the electron, respectively.
Here, the electron motion is along the $z$ direction.
By cycle-averaging the second term on the right-hand side, the ponderomotive potential is achieved, which provides the electron phase modulation effect:
\begin{eqnarray}
    U(\mathbf{R})=
        \frac{
            e^{2}\lambda^{2}I(\mathbf{R})
        }{
            8\pi^{2}m\varepsilon_{0}c^{3}
        },\label{eq:U}
\end{eqnarray}
where $\varepsilon_{0}$ represents the vacuum permittivity and $c$, $\lambda$ and $I$ denote the speed, wavelength and intensity of the light, respectively.
The non-relativistic case ($\gamma\sim1$) is considered herein.
The imprinted phase profile is equal to the intensity distribution of the light, and the modulation depth corresponds to the optical power.
In a milestone experiment performed by Schwartz et al., the intensity distribution of a laser standing wave resonating in an enhanced cavity was clearly visible as an electron phase-contrast image \cite{Schwartz2019}.
It is interesting to note that Bartell et al. and Schwarza et al. performed similar experiments using ruby- and neodymium-doped lasers soon after the laser was invented \cite{Bartell1965,Schwarza1965,Bartell1968}.
In these cases, the light beams intersected at right angles with respect to the electron beam; thus, the imprinted phase profile exhibited a two-fold rotational symmetry.
However, a circularly symmetric profile is preferable for practical electron microscopic applications. 
The lens action and SA correction can be important electron optical functionalities of such circular phase elements.
Recently, a structured light beam coaxially focused with an electron beam has been theoretically and numerically shown to act as a concave lens or produce a negative SA coefficient \cite{GarciadeAbajo2021,Uesugi2021}, which cannot be achieved using conventional electron round lenses.
Herein, a lens formed by the ponderomotive potential is referred to as a \textit{ponderomotive lens} (p-lens), just as an electron lens formed by the electrostatic potential is referred to as an electrostatic lens.
Figure \ref{fig:concept} shows a conceptual diagram of the p-lens discussed herein.
The incident electrons are parallel to a focused light beam propagating along the $z$ direction and are concentrated in the optical axis vicinity.

Using modern laser technology, the spatial distribution of the phase and polarisation of a light beam can be moderately modulated.
Hence, the parameters of a p-lens are expected to show considerably more degrees of freedom in design than those of electrostatic/magnetic lenses.
Before establishing a methodology for numerically designing a p-lens and the relevant structured light field, formulas are needed to guide the lens system design for the rapid implementation of p-lenses in electron microscopy and their effective verification.
Thus, this study aims to provide simple expressions for the properties of p-lenses derived from two familiar light beams: Bessel and Laguerre--Gaussian (LG) beams.
Based on the variational principle, the principles related to the lens properties in geometrical optics and the specific formulas in thin-lens approximations are presented.
Moreover, the results of electron trajectory calculations obtained using a p-lens in a lens system modelled from a scanning electron microscope are provided and the applicability of the p-lens as an electron lens and/or SA corrector is discussed.
Lastly, the discussions and conclusions are presented.

\section{Ray equation and properties of thin-lens}
The trajectory of an electron moving in a ponderomotive potential is derived using the principle of least action.
In an adiabatic process where the electron energy is conserved, the abbreviated action---also known as eikonal---is expressed as \cite{Landau}
\begin{eqnarray}
    S=\int_{A}^{B}\mathbf{p}\cdot d\mathbf{R},\label{eq:S}
\end{eqnarray}
where $\mathbf{p}$ denotes the canonical momentum.
When considering the motion of an electron in an electromagnetic field, the canonical momentum is usually replaced with $\mathbf{p}\to\mathbf{p}-e\mathbf{A}$.
However, in the present case, only the ponderomotive potential is considered because of the cycle-averaged light field in charge-free space.
Hence, the canonical momentum and the Hamiltonian of the electron are expressed as $\mathbf{p}=\gamma m\mathbf{v}$ and $H=\gamma mc^{2}+U$, respectively.
When the polarisation state is assumed to be perpendicular to the electron beam axis ($A_{z}=0$) or when handling non-relativistic electrons ($\gamma=1$), the ponderomotive potential is polarisation independent based on equation (\ref{eq:interactionH}).
Hence, the motion of electrons in a scalar potential (equation (\ref{eq:U})) is discussed.
The study focus is restricted to the circularly symmetric potential $U=U(r,z)$, where $r=\sqrt{x^{2}+y^{2}}$.

Rewriting the action into an expression using $z$ as the variable of integration, we obtain
\begin{eqnarray}
    S=\int_{A}^{B}p(x,y;z)\sqrt{1+x'^{2}+y'^{2}}\,dz,\label{eq:Sbyz}
\end{eqnarray}
where the prime symbol denotes the first derivative with respect to $z$ (i.e. $x'=\partial x/\partial z$).
The canonical momentum $p$---in which the electron velocity $\mathbf{v}$ is eliminated from the expression using the energy conservation $E=mc^{2}+T_{0}$---is expressed as
\begin{eqnarray}
    p=\sqrt{2m\tilde{U}},\label{eq:p}
\end{eqnarray}
where $\tilde{U}=(T_{0}-U)\left(1+T_{0}/2mc^{2}\right)$ and $T_{0}$ represents the initial kinetic energy of the electron.
Herein, the integrand on equation (\ref{eq:Sbyz}) is referred to as as the variational function:
\begin{eqnarray}
    F(x,x',y,y';z)=p\sqrt{1+x'^{2}+y'^{2}}.\label{eq:F}
\end{eqnarray}
The ray equation for deriving the electron trajectories is obtained by evaluating the variation of the action as $\delta S=0$.
For example, the expression for $x$ is
\begin{eqnarray}
    \frac{d}{dz}\frac{\partial F}{\partial x'}-\frac{\partial F}{\partial x}=0.\label{eq:rayeq}
\end{eqnarray}

When considering electrons whose trajectories are concentrated in the central axis vicinity of the potential, the variational function can be serially expanded for $x,x'$ and $y,y'$ as
\begin{eqnarray}
    F=F_{0}+F_{2}+F_{4}+F_{6}+\cdots,
\end{eqnarray}
where
\begin{eqnarray}
    \fl F_{0}=p_{0},\\
    \fl F_{2}=p_{2}(x^{2}+y^{2})+\frac{p_{0}}{2}(x'^{2}+y'^{2}),\\
    \fl F_{4}=p_{4}(x^{2}+y^{2})^{2}
        +\frac{p_{2}}{2}(x^{2}+y^{2})(x'^{2}+y'^{2})\nonumber\\
        -\frac{p_{0}}{8}(x'^{2}+y'^{2})^{2},\\
    \fl F_{6}=p_{6}(x^{2}+y^{2})^{3}
        +\frac{p_{4}}{2}(x^{2}+y^{2})^{2}(x'^{2}+y'^{2})\nonumber\\
        -\frac{p_{2}}{8}(x^{2}+y^{2})(x'^{2}+y'^{2})^{2}
        +\frac{p_{0}}{16}(x'^{2}+y'^{2})^{3},\label{eq:F6}
\end{eqnarray}
and $p_{n}=p_{n}(z)$ is the $n$-th coefficient in the $p$ expansion, which only has even-order terms because of the circular symmetry.
The variational function expanded to the $n$-th-order is expanded as $F^{(n)}=\sum_{i=0}^{n} F_{i}$, where the odd-order terms are considered to be 0.
The 0-th-order term represents the electron motion along the optical axis and contributes only to the acceleration or deceleration of the electron.
The magnitude of the ponderomotive potential that provides the lens effect (e.g. $O$(meV)) is negligible when compared with the kinetic energy of the electron (e.g. $O$(keV)).
Thus, the 0-th-order term can be replaced by the electron momentum in free space
\begin{eqnarray}
    F_{0}=p_{0}=\sqrt{2mT_{0}}.
\end{eqnarray}
The paraxial ray equation that provides the lens properties in Gaussian optics is derived by substituting $F^{(2)}$ in equation (\ref{eq:rayeq}):
\begin{eqnarray}
    x''-\frac{2p_{2}}{p_{0}}x=0.\label{eq:pararay}
\end{eqnarray}
The ray matrix can be obtained by adapting the thin-lens approximation, in which the distance of the electrons from the optical axis does not change before/after they enter the potential:
\begin{eqnarray}
    \left(
        \begin{array}{cc}
        x(B) \\
        x'(B) \\
        \end{array}
    \right)
    =
    \left(
        \begin{array}{cc}
        1 & 0 \\
        -1/f & 1 \\
        \end{array}
    \right)
    \left(
        \begin{array}{c}
        x(A) \\
        x'(A) \\
        \end{array}
    \right),
\end{eqnarray}
where $f$ denotes the focal length.
The focusing power is expressed as
\begin{eqnarray}
    \frac{1}{f}=-\frac{2}{p_{0}}\int_{-L/2}^{L/2}p_{2}(\xi)\,d\xi,\label{eq:f}
\end{eqnarray}
where $L$ denotes the interaction length between the electron and the potential and $\xi$ represents an axial coordinate with respect to the centre of the potential.
In the same approximation, the ray aberrations caused by the third- and fifth-order SAs can be expressed using simple formulas.
Aberrations in electron optics, particularly transmission electron microscopy, are generally defined at the object plane of a lens system.
Such aberrations are equal to the aberration at the image plane multiplied by the reciprocal of the transverse magnification of the lens system.
The third- and fifth-order SAs and the related SA coefficients are expressed as follows (refer to the Appendix for the derivation):
\begin{eqnarray}
    \Delta r_{\mathrm{S3}}=C_{\mathrm{S3}}\alpha_{o}^{3},\label{eq:raycs3}\\
    \Delta r_{\mathrm{S5}}=C_{\mathrm{S5}}\alpha_{o}^{5},\label{eq:raycs5}\\
    C_{\mathrm{S3}}=-\frac{4a^{4}}{p_{0}}\int_{-L/2}^{L/2}p_{4}(\xi)\,d\xi,\label{eq:cs3}\\
    C_{\mathrm{S5}}=-\frac{6a^{6}}{p_{0}}\int_{-L/2}^{L/2}p_{6}(\xi)\,d\xi,\label{eq:cs5}
\end{eqnarray}
where $\alpha_{o}$ represents an opening half-angle on the object plane side and $a$ represents the distance between an object and an entrance pupil of the lens.

\section{Properties of ponderomotive lenses}
First, the properties of the lens formed using a Bessel beam is investigated, which represents a solution of the wave equation and is appropriate for describing a tightly focused light beam at a high numerical aperture with a prominent longitudinal component of the electric or magnetic field.
A scalar Bessel beam in the cylindrical coordinate system is expressed as \cite{Durnin1987,Palma1996}
\begin{eqnarray}
    \psi_{\mathrm{B}}(\mathbf{R})=J_{n}(K_{0}r)\exp{(in\phi)}\exp{\left(iz\sqrt{k^{2}-K_{0}^{2}}\right)},
\end{eqnarray}
where $J_{n}$ denotes the Bessel function of the first kind of order $n$, $K_{0}=k\sin\theta_{0}$, $k=2\pi/\lambda$ and $\theta_{0}$ denotes the cone half-angle of the beam convergence.
The azimuthal phase term $in\phi$ indicates that this beam carries angular momentum; hence the beam is known as a type of optical vortex.
Using this expression, the ponderomotive potential is expresed as
\begin{eqnarray}
    U(r)=U_{0}J_{n}^{2}(K_{0}r),
    \label{eq:BesselU}
\end{eqnarray}
where $U_{0}$ denotes the magnitude of the potential properly associated with the light intensity using equation (\ref{eq:U}).
The potential is independent of $z$ because of the nondiffracting properties of the Bessel beam.

Substituting equations (\ref{eq:BesselU}) into (\ref{eq:p}) and expanding yields the expansion coefficients of $p$ and the lens properties.
Table \ref{tbl:bessel} summarises $p_{n}$ and the lens properties obtained using the $J_{n}$ Bessel beam up to $n=3$.
Herein, the following relation that holds in the non-relativistic case is assumed:
\begin{eqnarray}
    U_{0}\ll T_{0}\ll mc^{2}.\label{eq:appro}
\end{eqnarray}
The $J_{0}$ and $J_{1}$ Bessel beams can function as concave and convex lenses, respectively, while the others do not produce the focusing power.
Regarding SA coefficients, the $J_{1}$ case showed a negative $C_{\mathrm{S3}}$, whereas the $J_{1}$ case with $n\geq2$ showed no third-order SA.
The $J_{1}$ Bessel beam can be used to correct the third-order SA of an electron lens system; however, it worsens the fifth-order SA because of its positive $C_{\mathrm{S5}}$.
\begin{table}[tbp]
    \caption{
        \label{tbl:bessel}
        Series expansion coefficients of the canonical momentum, focusing power and third- and fifth-order spherical aberration coefficients of ponderomotive lenses using a Bessel beam of order $n$.
    }
    \begin{indented}
        \item[]
        \begin{tabular}{@{}lllll}
            \br
$n$
    &0
        &1
            &2
                &3\\
            \mr
$p_{2}$
    &$\frac{p_{0}U_{0}K_{0}^{2}}{4T_{0}}$
        &$\frac{-p_{0}U_{0}K_{0}^{2}}{8T_{0}}$
            &0
                &0\\
$p_{4}$
    &$\frac{-3p_{0}U_{0}K_{0}^{4}}{64T_{0}}$
        &$\frac{p_{0}U_{0}K_{0}^{4}}{32T_{0}}$
            &$\frac{-p_{0}U_{0}K_{0}^{4}}{128T_{0}}$
                &0\\
$p_{6}$
    &$\frac{5p_{0}U_{0}K_{0}^{6}}{1152T_{0}}$
        &$\frac{-5p_{0}U_{0}K_{0}^{6}}{1536T_{0}}$
            &$\frac{p_{0}U_{0}K_{0}^{6}}{768T_{0}}$
                &$\frac{-p_{0}U_{0}K_{0}^{6}}{4608T_{0}}$\\
$1/f$
    &$\frac{-U_{0}K_{0}^{2}L}{2T_{0}}$
        &$\frac{U_{0}K_{0}^{2}L}{4T_{0}}$
            &0
                &0\\
$C_{\mathrm{S3}}$
    &$\frac{3U_{0}K_{0}^{4}a^{4}L}{16T_{0}}$
        &$\frac{-U_{0}K_{0}^{4}a^{4}L}{8T_{0}}$
            &$\frac{U_{0}K_{0}^{4}a^{4}L}{32T_{0}}$
                &0\\
$C_{\mathrm{S5}}$
    &$\frac{-5U_{0}K_{0}^{6}a^{6}L}{192T_{0}}$
        &$\frac{5U_{0}K_{0}^{6}a^{6}L}{256T_{0}}$
            &$\frac{-U_{0}K_{0}^{6}a^{6}L}{128T_{0}}$
                &$\frac{U_{0}K_{0}^{6}a^{6}L}{768T_{0}}$\\
            \br
        \end{tabular}
    \end{indented}
\end{table}

Next, the lens properties of the p-lenses are shown using LG beams.
LG beams are solutions of the paraxial Helmholtz equation with a sufficiently smaller convergence/divergence angle than unity.
Although Bessel--Gaussian beams, which are paraxial versions of Bessel beams, were available \cite{Gori1987,Borghi2001}, LG beams were selected for the investigation because they are more common in laser optics and exhibit circular symmetry profiles.
The formulas for the Bessel-Gaussian beams can be derived in the same manner as those for the LG beams and show similar characteristics; however, the expressions are slightly complicated.

A scalar LG beam is expressed as \cite{Yariv}
\begin{eqnarray}
    \fl\psi_{\mathrm{LG}}(\mathbf{R})=\nonumber\\
        \frac{w_{0}}{w(z)}
        \left(\frac{\sqrt{2}r}{w(z)}\right)^{|\ell|}
        L^{|\ell|}_{p}
        \left(\frac{2r^{2}}{w^{2}(z)}\right)
        \exp\left(
            -\frac{r^{2}}{w^{2}(z)}
        \right)\nonumber\\
        \exp\left[
            \frac{ikr^{2}}{2R(z)}
            +i\ell\phi
            +i(2p+|\ell|+1)\eta(z)
        \right],
        \label{eq:LG}
\end{eqnarray}
with
\begin{eqnarray}
    w(z)=w_{0}\sqrt{1+z^{2}/z_{\mathrm{R}}^{2}},\\
    R(z)=z(1+z_{\mathrm{R}}^{2}/z^{2}),\\
    \eta(z)=\tan^{-1}(z/z_{\mathrm{R}}),
\end{eqnarray}
where $w_{0}$ represents the Gaussian waist radius, $z_{\mathrm{R}}=kw_{0}^{2}/2$ denotes the Rayleigh length, $L^{|\ell|}_{p}$ denotes the Laguerre polynomials and $p$ and $\ell$ represent the radial and azimuthal indices, respectively.
Equation (\ref{eq:LG}) yields the fundamental Gaussian mode in the case of $L_{0}^{0}$, whereas for non-zero $\ell$, the LG beam exhibits an azimuthal phase term and thus is an optical vortex.
The ponderomotive potential obtained using the $L_{0}^{\ell}$ LG beam is expressed as
\begin{eqnarray}
    U(r,z)=
        \frac{U_{0}w_{0}^{2}}{w(z)^{2}}
        \left(
            \frac{\sqrt{2}r}{w(z)}
        \right)^{2|\ell|}
        \exp\left(
            -\frac{2r^{2}}{w^{2}(z)}
        \right).\label{eq:LGU}
\end{eqnarray}

Unlike the Bessel beam, the obtained potential shows a $z$-dependence.
The integral is performed over $\xi$ in equations (\ref{eq:f}), (\ref{eq:cs3}) and (\ref{eq:cs5}).
The potential obtained using equation (\ref{eq:LGU}) is localised within a few times of $z_{\mathrm{R}}$ around the beam waist; hence, $z_{\mathrm{R}}$ is assumed to be small when compared with the scale of the lens system (i.e. $L\ll z_{\mathrm{R}}$) in the thin-lens approximation.
Table \ref{tbl:LG} summarises the expansion coefficients of $p$ and the lens properties for the $L^{\ell}_{0}$ LG beam up to $\ell=3$.
The p-lenses formed using the LG beams exhibit properties similar to those obtained using the Bessel beams.
The $L^{1}_{0}$ LG beam is the only beam with a negative $C_{\mathrm{S3}}$.
Such characteristics are true for LG beams with non-zero $p$ because a radial index increment of unity only adds an outermost annulus to the beam profile.
\begin{table}
    \caption{
        \label{tbl:LG}
        Series expansion coefficients of the canonical momentum, focusing power and third- and fifth-order spherical aberration coefficients of ponderomotive lenses using a Laguerre--Gaussian beam with $L^{\ell}_{0}$.
    }
    \begin{indented}
        \item[]
        \begin{tabular}{@{}lllll}
            \br
$\ell$
    &0
        &1
            &2
                &3\\
            \mr
$p_{2}$
    &$\frac{p_{0}U_{0}w_{0}^{2}}{T_{0}w^{4}}$
        &$\frac{-p_{0}U_{0}w_{0}^{2}}{T_{0}w^{4}}$
            &0
                &0\\
$p_{4}$
    &$\frac{-p_{0}U_{0}w_{0}^{2}}{T_{0}w^{6}}$
        &$\frac{2p_{0}U_{0}w_{0}^{2}}{T_{0}w^{6}}$
            &$\frac{-2p_{0}U_{0}w_{0}^{2}}{T_{0}w^{6}}$
                &0\\
$p_{6}$
    &$\frac{2p_{0}U_{0}w_{0}^{2}}{3T_{0}w^{8}}$
        &$\frac{-2p_{0}U_{0}w_{0}^{2}}{T_{0}w^{8}}$
            &$\frac{4p_{0}U_{0}w_{0}^{2}}{T_{0}w^{8}}$
                &$\frac{-4p_{0}U_{0}w_{0}^{2}}{T_{0}w^{8}}$\\
$1/f$
    &$\frac{-\pi U_{0}z_{\mathrm{R}}}{T_{0}w_{0}^{2}}$
        &$\frac{\pi U_{0}z_{\mathrm{R}}}{T_{0}w_{0}^{2}}$
            &0
                &0\\
$C_{\mathrm{S3}}$
    &$\frac{3\pi U_{0}a^{4}z_{\mathrm{R}}}{2T_{0}w_{0}^{4}}$
        &$\frac{-3\pi U_{0}a^{4}z_{\mathrm{R}}}{T_{0}w_{0}^{4}}$
            &$\frac{3\pi U_{0}a^{4}z_{\mathrm{R}}}{T_{0}w_{0}^{4}}$
                &0\\
$C_{\mathrm{S5}}$
    &$\frac{-5\pi U_{0}a^{6}z_{\mathrm{R}}}{4T_{0}w_{0}^{6}}$
        &$\frac{15\pi U_{0}a^{6}z_{\mathrm{R}}}{4T_{0}w_{0}^{6}}$
            &$\frac{-15\pi U_{0}a^{6}z_{\mathrm{R}}}{2T_{0}w_{0}^{6}}$
                &$\frac{15\pi U_{0}a^{6}z_{\mathrm{R}}}{2T_{0}w_{0}^{6}}$\\
            \br
        \end{tabular}
    \end{indented}
\end{table}

\section{Spherical aberration correction using ponderomotive lenses}
Using the aforementioned formulas, a p-lens is designed and its applicability is evaluated.
Considering the need to estimate the required optical power subsequently, the LG beam is primarily discussed in this section.
The Bessel beam possesses infinite energy and does not exist in reality.
Each formula for the LG beam is rewritten as an expression that is not explicit in $T_{0}$ and $U_{0}$, as shown in table \ref{tbl:fLG}, where $\mathcal{F}$ is a length parameter expressed using $1/\mathcal{F}=-\pi U_{0}z_{\mathrm{R}}/T_{0}w_{0}^{2}$ and $1/\mathcal{F}=\pi U_{0}z_{\mathrm{R}}/T_{0}w_{0}^{2}$ for $\ell=2$ and $3$, respectively.

\begin{figure}[tbp]
    \centering
    \includegraphics[width=73mm]{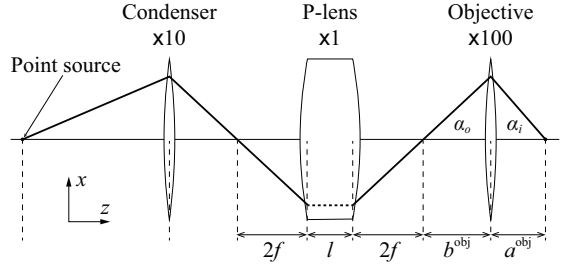}
    \caption{
        Schematic of a lens system in the electron trajectory calculation.
        A scanning electron microscopic system is assumed.
        A point electron source is along the electron optical axis.
        Condenser and objective electron lenses are thin lenses.
        Table \ref{tbl:para} shows the objective lens properties.
        A p-lens with a thickness of $l$ acts as a relay lens with unity magnification.
    }
    \label{fig:schem}
\end{figure}

Figure \ref{fig:schem} shows the lens system used in the electron trajectory calculation, which comprises a point electron source placed on an electron beam axis, a thin condenser lens with a magnification of 10, a p-lens produced using the $L_{0}^{1}$ LG beam with unity magnification and a thin objective lens with a typical third-order SA.
Table \ref{tbl:para} presents the parameters of the objective lens.
The refractive index distribution of the p-lens is consistent with the intensity distribution of the LG beam.
Moreover, the p-lens shows a thickness of $l$, which is the distance between the primary and secondary principal planes.
Herein, a scanning electron microscopic system is assumed as the lens system.
Consequently, the opening angle $\alpha_{o}$ associated with the ray aberrations (equations (\ref{eq:raycs3}) and (\ref{eq:raycs5})) is replaced with $\alpha_{i}$, which represents an opening half-angle at the image plane or the objective lens focus.
Accordingly, $a$ is also redefined as the distance between the lens exit pupil and the image plane.
\begin{table}
    \caption{
        \label{tbl:fLG}
        Formulas in table \ref{tbl:LG} rewritten using $f$ or $\mathcal{F}$.
    }
    \begin{indented}
        \item[]
        \begin{tabular}{@{}lllll}
            \br
$\ell$
    &0
        &1
            &2
                &3\\
            \mr
$p_{2}$
    &$\frac{-p_{0}w_{0}^{4}}{\pi fz_{\mathrm{R}}w^{4}}$
        &$\frac{-p_{0}w_{0}^{4}}{\pi fz_{\mathrm{R}}w^{4}}$
            &0
                &0\\
$p_{4}$
    &$\frac{p_{0}w_{0}^{4}}{\pi fz_{\mathrm{R}}w^{6}}$
        &$\frac{2p_{0}w_{0}^{4}}{\pi fz_{\mathrm{R}}w^{6}}$
            &$\frac{2p_{0}w_{0}^{4}}{\pi \mathcal{F}z_{\mathrm{R}}w^{6}}$
                &0\\
$p_{6}$
    &$\frac{-2p_{0}w_{0}^{4}}{3\pi fz_{\mathrm{R}}w^{8}}$
        &$\frac{-2p_{0}w_{0}^{4}}{\pi fz_{\mathrm{R}}w^{8}}$
            &$\frac{-4p_{0}w_{0}^{4}}{\pi \mathcal{F}z_{\mathrm{R}}w^{8}}$
                &$\frac{-4p_{0}w_{0}^{4}}{\pi \mathcal{F}z_{\mathrm{R}}w^{8}}$\\
$C_{\mathrm{S3}}$
    &$\frac{-3a^{4}}{2fw_{0}^{2}}$
        &$\frac{-3a^{4}}{fw_{0}^{2}}$
            &$\frac{-3a^{4}}{\mathcal{F}w_{0}^{2}}$
                &0\\
$C_{\mathrm{S5}}$
    &$\frac{5a^{6}}{4fw_{0}^{4}}$
        &$\frac{15a^{6}}{4fw_{0}^{4}}$
            &$\frac{15a^{6}}{2\mathcal{F}w_{0}^{4}}$
                &$\frac{15a^{6}}{2\mathcal{F}w_{0}^{4}}$\\
            \br
        \end{tabular}
    \end{indented}
\end{table}
\begin{table}
    \caption{
        \label{tbl:para}
        Parameters of an objective lens with positive third-order spherical aberration are used in the electron trajectory calculation.
    }
    \begin{indented}
        \item[]
        \begin{tabular}{@{}llll}
            \br
Parameter&
    Symbol&
        Value&
            Unit\\
            \mr
Focal length&
    $f^{\mathrm{obj}}$&
        1&
            mm\\
Magnifying power&
    $\beta$&
        100&
            n/a\\
Object distance&
    $a^{\mathrm{obj}}$&
        1.01&
            mm\\
Image distance&
    $b^{\mathrm{obj}}$&
        101&
            mm\\
Object opening half-angle&
    $\alpha_{o}$&
        0.1&
            mrad\\
Image opening half-angle&
    $\alpha_{i}$&
        10&
            mrad\\
Third-order SA coefficient&
    $C_{\mathrm{S3}}^{\mathrm{obj}}$&
        1&
            mm\\
Third-order SA at $\alpha_{i}$&
    $\Delta r_{\mathrm{S3}}^{\mathrm{obj}}$&
        1&
            nm\\
            \br
        \end{tabular}
    \end{indented}
\end{table}

The $\Delta r_{\mathrm{S3}}$ magnitude required for a p-lens to correct the SA of the objective lens is $1\times\beta=100$ nm, while its opening half-angle is $10/\beta=0.1$ mrad.
Thus, the target value for the third-order SA coefficient is $-10^{-7}/(10^{-4})^{3}=-10^{5}$ m.
Using the $C_{\mathrm{S3}}$ formula in table \ref{tbl:fLG} and the relation $a=2f$, the focal length expression is obtained as a function of $w_{0}$: $f=(w_{0}^{2}/48\times10^{5})^{1/3}$ m.
For the two cases of $w_{0}=10\lambda$ and $100\lambda$ with $\lambda=1$ $\mathrm{\mu m}$, $f=5.93$ and $27.5$ mm, respectively, are obtained.

\begin{figure}[tbp]
    \centering
    \includegraphics[width=73mm]{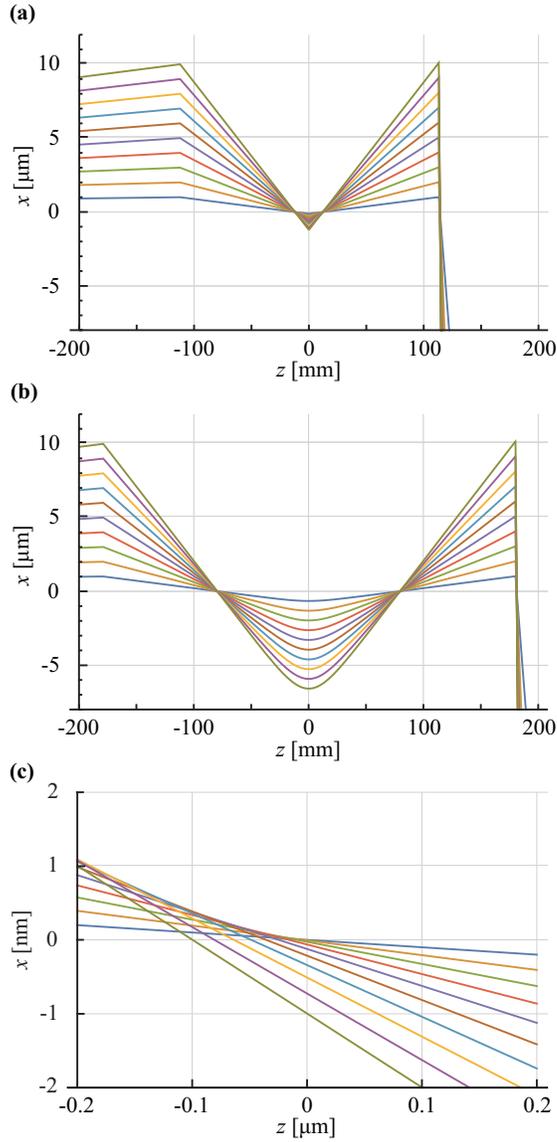}
    \caption{
        Electron trajectories around the ponderomotive lenses obtained using the $L_{0}^{1}$ LG beam by employing the variational function expanded to the second order.
        (a) Case of $w_{0}=10\lambda$.
        (b) Case of $w_{0}=100\lambda$.
        (c) Enlarged image of the focal area of an electron objective lens with $C_{\mathrm{S3}}^{\mathrm{obj}}=1$ mm. $z=0$ is set at the centre of the ponderomotive lens in (a) and (b) and at the focus in (c).
    }
    \label{fig:F2}
\end{figure}

First, the electron trajectory with a p-lens variational function, which is expanded to the second order, is calculated to determine the optimal $l$ value.
The obtained focal length is introduced under the thin-lens approximation; thus, if $l$ is not optimal, the p-lens will not function as a relay lens with unity magnification.
Figures \ref{fig:F2}(a) and (b) show the calculation results with $l=0.94z_{\mathrm{R}}$ and $0.765z_{\mathrm{R}}$ for $w_{0}=10\lambda$ and $100\lambda$, respectively.
Using these values, the p-lenses function as the relay lens. In both cases, the results show the ray aberration of 1 nm at the focus of the objective lens (figure \ref{fig:F2}(c)).
The p-lenses are Gaussian lenses; hence, this aberration is caused only by the objective lens.

\begin{figure}[tbp]
    \centering
    \includegraphics[width=73mm]{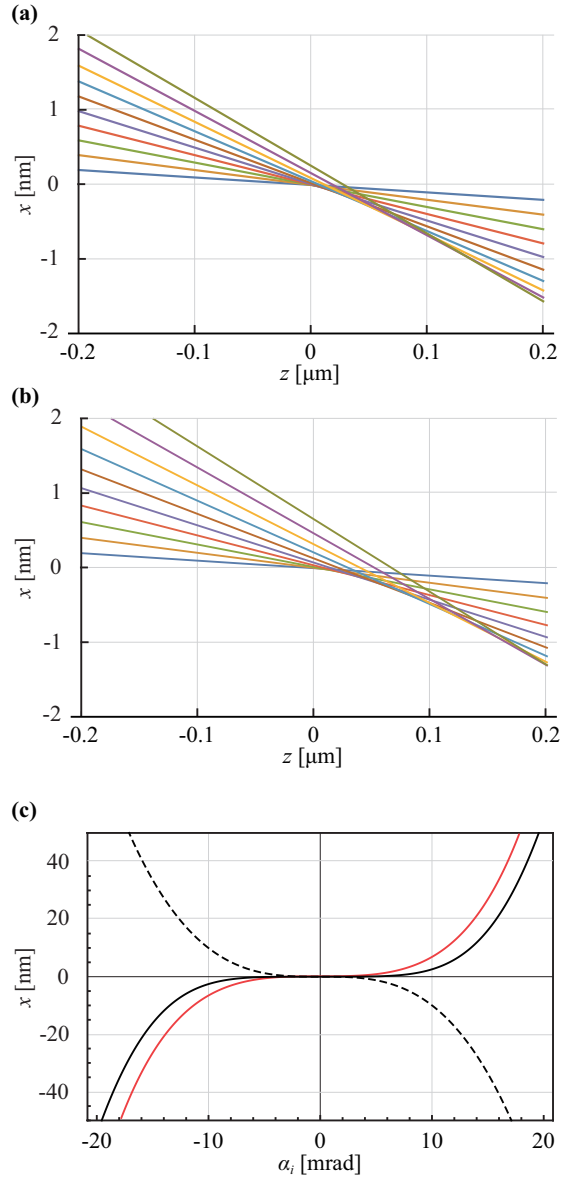}
    \caption{
        Electron trajectories at the focus of an objective lens calculated using the full variational function. (a) Case of $w_{0}=10\lambda$. (b) Case of $w_{0}=100\lambda$.
        (c) Transverse aberration diagram.
        $\alpha_{i}$ represents an opening half-angle at the focus.
        Black and red solid curves represent the cases of $10\lambda$ and $100\lambda$, respectively.
        The dashed curve represents for the objective lens aberration.
    }
    \label{fig:Fw10and100}
\end{figure}

Subsequently, the electron trajectory calculation with a p-lens using the full variational function is performed.
Both $T_{0}$ and $U_{0}$ can be eliminated from the expression of $F$ by substituting the formula for $f$.
Figures \ref{fig:Fw10and100}(a) and (b) show the electron trajectories at the focus for $w_{0}=10\lambda$ and $100\lambda$, respectively.
The positive SA of the objective lens is corrected using the p-lenses.
However, some overcompensations exist and the error is larger when $w_{0}=100\lambda$ than when $w_{0}=10\lambda$ because the $z$ distribution of the potential became larger as $w_{0}$ increased.
Figure \ref{fig:Fw10and100}(c) shows the aberration diagram, where $\alpha_{i}$ denotes the opening half-angle of the source converted to the angle at the objective lens focus.
The black and red solid curves represent the cases of $10\lambda$ and $100\lambda$, respectively.
The dashed curve denotes the objective lens aberration.
The diagram shows that the SA is corrected up to $\sim7$ mrad for $w_{0}=10\lambda$ but only up to $\sim4$ mrad for $w_{0}=100\lambda$.

\begin{figure}[tbp]
    \centering
    \includegraphics[width=73mm]{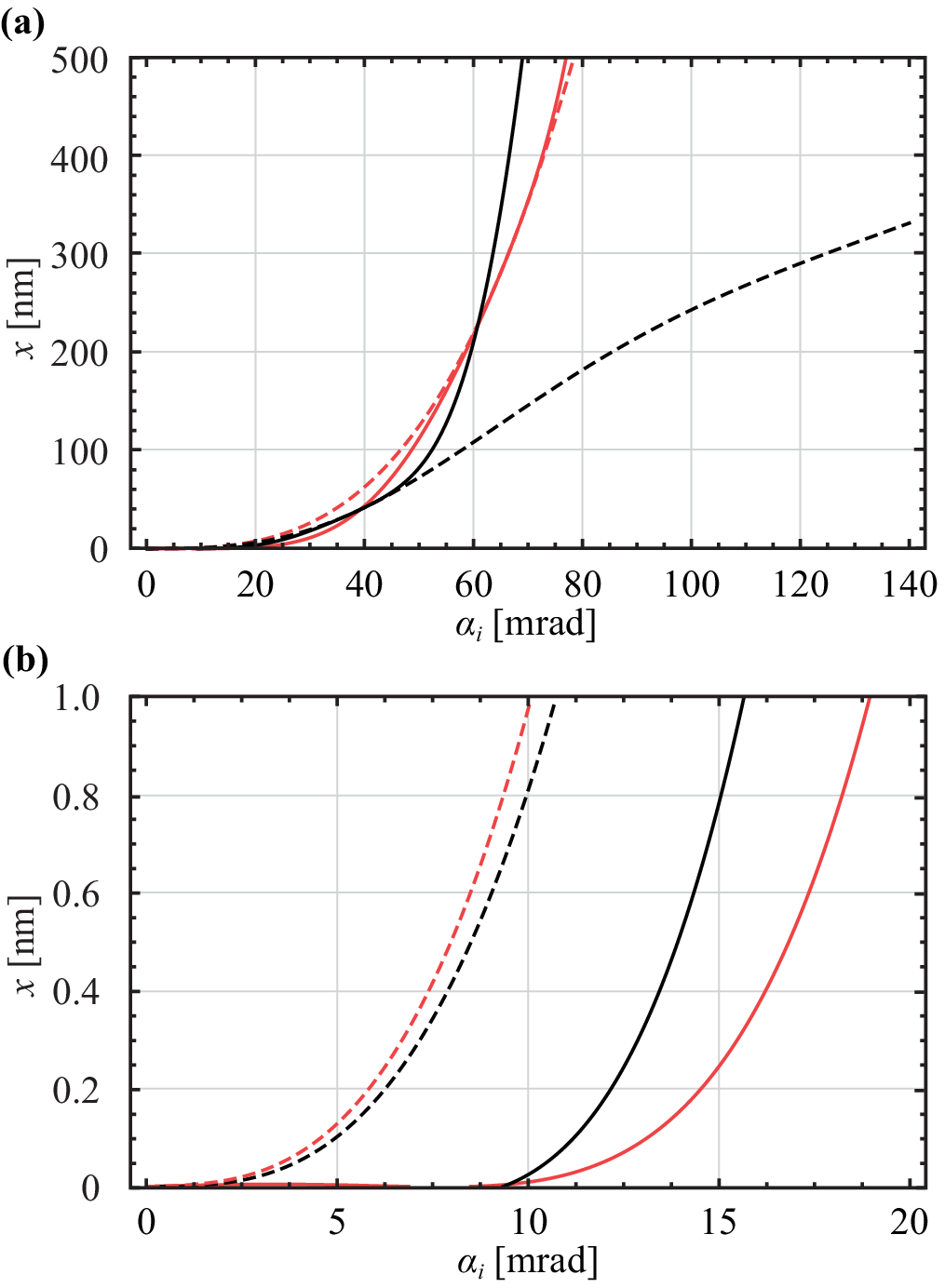}
    \caption{
        Transverse aberration diagrams with the optimal focal length $f$.
        $\alpha_{i}$ represents an opening half-angle at the focus.
        Solid curves represent the entire lens systems.
        Dashed curves represent ponderomotive lenses.
        Black an red curves denote the case of $w_{0}=10\lambda$ and $100\lambda$, respectively.
        (a) Range up to 20 mrad.
        (b) Range up to 140 mrad.
    }
    \label{fig:Fw10and100_optimum}
\end{figure}

By using $f$ derived from the formula as a guideline for lens designing, the optimal $f$ value and the corresponding $l$ value for minimising the lens system aberration within $\alpha_{i}\leq10$ mrad are determined.
The obtained $f$ and $l$ values are 5.5 mm and $0.94z_{\mathrm{R}}$, respectively, for $w_{0}=10\lambda$ and 22.3 mm and $0.7253z_{\mathrm{R}}$, respectively, for $w_{0}=100\lambda$.
Figure \ref{fig:Fw10and100_optimum} shows the calculation results of the aberration diagrams using these parameters.
The solid curves depict the aberrations of the entire lens system.
The dashed curves represent aberrations when the objective lens has no third-order SA (i.e. $C_{\mathrm{S3}}^{\mathrm{obj}}=0$).
The black and red colours depict the cases of $10\lambda$ and $100\lambda$, respectively.
The third-order SA is well corrected up to 10 mrad in both cases (Figure \ref{fig:Fw10and100_optimum}(a)), whereas $100\lambda$ exhibits a smaller higher-order aberration up to 40 mrad.
The dashed curves represent the aberration curves of the two p-lenses.
Figure \ref{fig:Fw10and100_optimum}(b) shows that the curve shape for $w_{0}=10\lambda$ exhibits a ripple that corresponds to the annular profile of the LG beam.
The curve for $w_{0}=100\lambda$ also shows the same shape (not shown in the figure).
The difference between the two cases is the scaling determined using parameters $f$ and $w_{0}$.

\begin{figure}[tbp]
    \centering
    \includegraphics[width=73mm]{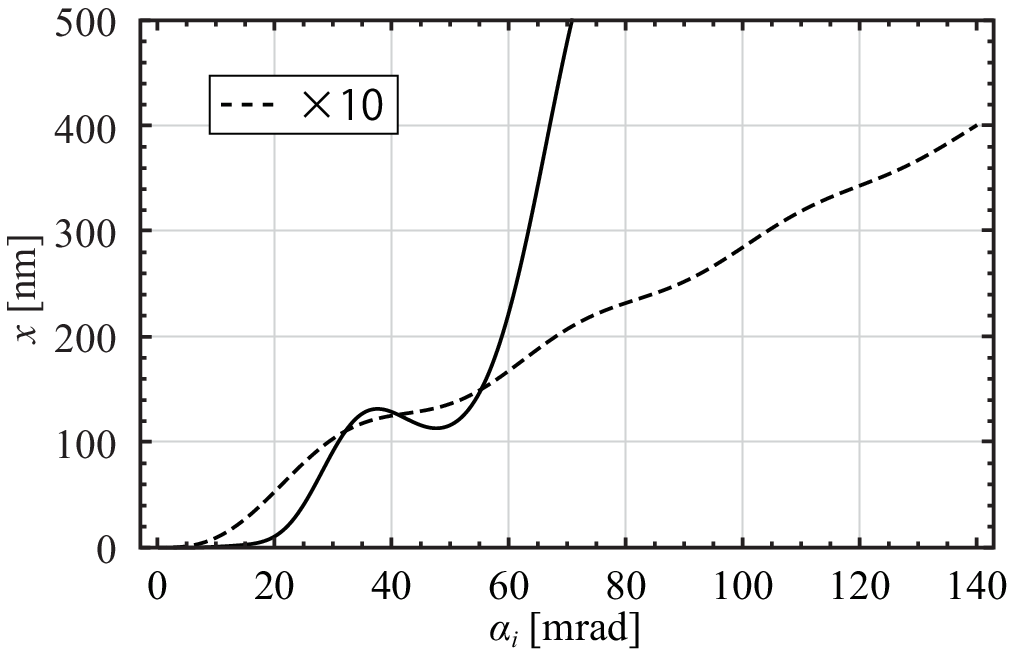}
    \caption{
        Transverse aberration diagram for a ponderomotive lens obtained using the $J_{1}$ Bessel beam.
        $\alpha_{i}$ represents an opening half-angle at the focus.
        Solid curve represents the entire lens system.
        Dashed curve denotes the aberration property of the ponderomotive lens and is magnified by a factor of 10.
    }
    \label{fig:Bessel}
\end{figure}

A similar ripple structure of the aberration curve is also observed in the case of the Bessel beam.
Figure \ref{fig:Bessel} shows the aberration diagram of the p-lens obtained using the $J_{1}$ Bessel beam.
The design parameters for correcting the SA of the objective lens are $\lambda=1$ $\mathrm{\mu m}$, $\theta_{0}=70^{\circ}$, $L=10\lambda$ and $f=2$ mm.
The solid curve represents the entire lens system.
The SA of the objective lens is corrected using the p-lens up to $\sim$5 mrad.
The dashed curve denotes the p-lens and is magnified by a factor of 10 for ease of viewing.
Multiple ripples appear on the aberration curve because of the multi-ring profile of the Bessel beam.
Compared with the results shown in figure \ref{fig:Fw10and100_optimum}, a larger higher-order SA occurs in the smaller opening-angle range owing to the smaller annular diameter of this Bessel beam compared to that of the LG beams.

\section{Discussion}
When applying p-lenses, it is important to determine whether the focal lengths obtained in the previous section are practical.
For the $L_{0}^{1}$ LG beam, the expression for the optical power derived from equations (\ref{eq:U}) and (\ref{eq:LGU}) yields
\begin{eqnarray}
    P=\frac{
        4\pi^{2}\varepsilon_{0}mc^{3}T_{0}w_{0}^{4}
    }{
        e^{2}\lambda^{2}z_{\mathrm{R}}f
    }.
\end{eqnarray}
When the electron energy is $T_{0}=1$ keV and the wavelength is $\lambda=1$ $\mathrm{\mu m}$, the required optical powers for $w_{0}=10\lambda$ and $100\lambda$ and $f=5.93$ and 27.5 mm are 287 kW and 6.20 MW, respectively.
Such optical power can be achieved using an ultrafast laser with a pulse energy in the $\mathrm{\mu J}$ order, which is readily available.
Furthermore, using an enhanced cavity, the $O(100)$-kW optical power can be obtained as an average power rather than a peak power \cite{Carstens2014}.
However, the technical difficulty of preparing a cavity with holes for an electron beam to pass through must be overcome to employ this approach.

The sign of the SA coefficient of the p-lenses obtained using the $J_{1}$ Bessel and $L_{0}^{1}$ LG beams is reversed whenever its order increases by two, affording an aberration curve with a unique rippled shape.
Alternatively, the overall shape of the aberration curve is monotonic and always produces a negative SA coefficient. Optimising an objective lens system to compensate for the overcompensation of the p-lens, for example, by combining some electrostatic/magnetic round lenses with different SAs may allow SA correction in a wide angular range. 
Because of the annular profile of the light beams and considering that the ponderomotive potential acts as a repulsive potential, the p-lenses obtained using the second- and higher-order Bessel and LG beams are also expected to constantly produce a negative SA coefficient in the overall aberration curve.
These beams seem to provide the best potential distribution for use as aberration correction plates because of no focusing power; however, no beam produces a negative third-order SA, limiting the application of such beams. 
Although Garc\'{i}a de Abajo et al. reported that the $L_{0}^{3}$ LG beam can be applied to the third-order SA correction \cite{GarciadeAbajo2021}, a similar result is not achieved in the present work.
Further theoretical studies and experimental verification are needed for the correction technique using the higher-order optical vortex.

The results obtained herein are mainly concerned with non-relativistic electrons.
The intensity distributions of the Bessel and LG beams show no dependence on the azimuthal phase; therefore, the angular momentum coupling between the electrons and optical vortex beams is not realised.
Alternatively, the spin--orbit interactions of photons and electrons occur in the relativistic regime \cite{Kaplan2005,Smorenburg2011,Ahrens2012}.
Coupling may occur if we consider the superposition of two or more optical vortex beams \cite{Handali2015,Kozak2021}.
The accumulated phase shift along the electron trajectory may be used as a helical phase plate by employing a twisted potential distribution generated by the difference in the Gouy phase of two coaxially superposed optical vortices \cite{Hamazaki2006,Droop2021}.

\section{Conclusion}
The properties of electron lenses based on the refractive index distribution obtained using ponderomotive potentials with scalar Bessel and LG light beams are investigated on the basis of the variational principle.
The simple formulas derived from the thin-lens approximation are appropriate for determining the focal length and SA of p-lenses, which are useful as guiding parameters in the lens design.
The calculation for solving the ray equation does not require information about the initial energy of an electron or the magnitude of the ponderomotive potential but rather demonstrates that the focal length is the most important parameter in the determination of the lens properties.

Only the $J_{1}$ Bessel or $L_{0}^{1}$ LG beams functioning as convex lenses exhibit a negative third-order SA and can be used for the SA correction of conventional electron lens systems, consistent with the results reported in the literature \cite{Uesugi2021}.
A smaller scaling parameter ($w_{0}$) in the case of the LG beam brings the actual lens properties closer to those obtained using the formulas, whereas a larger scaling parameter reduces the influence of a higher-order SA over a wider range of opening angles.
The beams of all orders, except for the 0-th-order LG beam (i.e. the Gaussian beam), are expected to exhibit curves with a ripple structure in an aberration diagram and to achieve an SA with a sign opposite to that of an electrostatic/magnetic lens.

For non-relativistic electrons, the ponderomotive potential is proportional to the light intensity.
The next topic in the development of p-lenses must be the design of the spatial amplitude and the phase distribution of a light beam to realise an arbitrary refractive index distribution.
The lens properties showed herein obtained using well-known light beams---such as Bessel and LG beams---will serve as a guideline for promoting such future research and advancing novel electron optics that exploit the electron--light interaction.

\ack
We would like to thank Prof. K. Saitoh for helpful discussions and comments on this work.
This work was supported by JSPS KAKENHI Grant Number JP20H02629, JST, PRESTO Grant Number JPMJPR2004, and the joint research programme of the Institute of Materials and Systems
for Sustainability, Nagoya University.

\appendix
\setcounter{section}{1}
\section*{Appendix}
\begin{figure}[tbp]
    \centering
    \includegraphics[width=73mm]{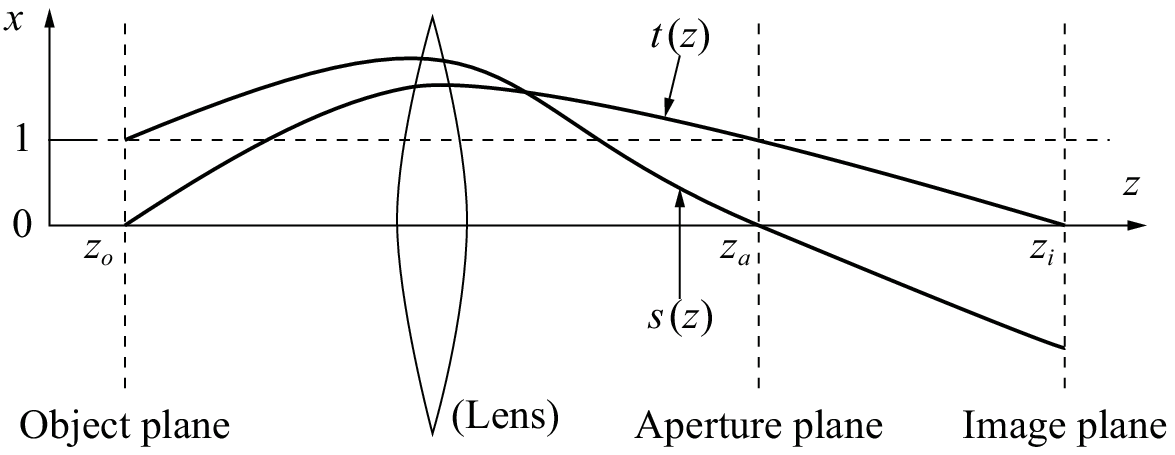}
    \caption{
        Schematic of the fundamental solutions of the ray equation following the boundary conditions in equation (\ref{eq:st}).
    }
    \label{fig:st}
\end{figure}

\begin{figure}[tbp]
    \centering
    \includegraphics[width=73mm]{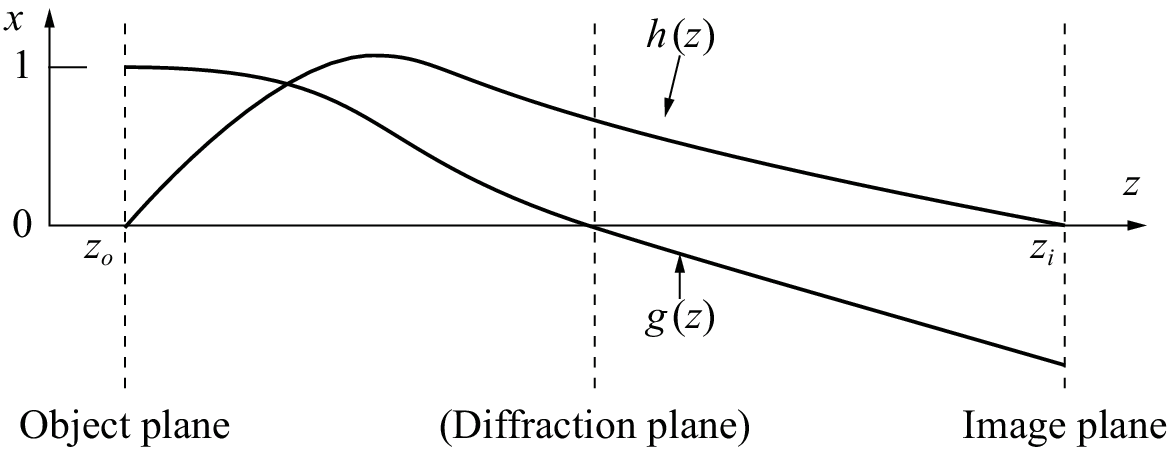}
    \caption{
        Schematic of the fundamental solutions of the ray equation following the boundary conditions in equation (\ref{eq:gh}).
    }
    \label{fig:gh}
\end{figure}

The derivation of SA coefficients is described here.
The fourth-order and subsequent expansion terms of the variational function show the deviation of an electron trajectory from the Gaussian trajectory, which is a solution of the paraxial ray equation (equation (\ref{eq:pararay})).
The fourth-order expansion term is rewritten as
\begin{eqnarray}
    \fl-F_{4}=
        \frac{L}{4}(x^{2}+y^{2})^{2}
        +\frac{M}{2}(x^{2}+y^{2})(x'^{2}+y'^{2})\nonumber\\
        +\frac{N}{4}(x'^{2}+y'^{2})^{2},\label{eq:F4}
\end{eqnarray}
where $L=-4p_{4},\,M=-p_{2},\,N=p_{0}/2$ are provided in accordance with the convention of electron optics \cite{Kato2006}.
The aforementioned expression is shown as a function of ray heights $x$ and $y$ and slopes $x'$ and $y'$.
The aberration at the image plane is also affected by the size and position of a lens aperture; therefore, the aberration function should be specified based on the trajectory between the image and aperture planes.
The fundamental solutions of the ray equation, $s(z)$ and $t(z)$, that satisfy the following boundary conditions at an object plane $z=z_{o}$ and aperture plane $z=z_{a}$ are considered \cite{Sturrock}:
\begin{eqnarray}
    \left\{
        \begin{array}{l}
            s(z_{o})=1\\
            s(z_{a})=0
        \end{array}
    \right.,\quad
    \left\{
        \begin{array}{l}
            t(z_{o})=0\\
            t(z_{a})=1
        \end{array}
    \right..\label{eq:st}
\end{eqnarray}
Figure \ref{fig:st} shows a schematic of these trajectories.
Although the aperture plane in the figure is defined downstream of the lens, it can be located anywhere between the object and image planes.
Based on the linear combination of these solutions, the general solutions of the ray equation for $x$ and $y$ can be obtained:
\begin{eqnarray}
    \left\{
        \begin{array}{l}
            x(z) =x_{o}s(z)+x_{a}t(z)\\
            x'(z) =x_{o}s'(z) +x_{a}t'(z)
        \end{array}
    \right.,\label{eq:xx}\\
    \left\{
        \begin{array}{l}
            y(z) =y_{o}s(z)+y_{a}t(z)\\
            y'(z) =y_{o}s'(z) +y_{a}t'(z)
        \end{array}
    \right.,\label{eq:yy}
\end{eqnarray}
where $x_{o}=x(z_{o})$, $x_{a}=x(z_{a})$, $y_{o}=y(z_{o})$ and $y_{a}=y(z_{a})$.
Substitute the solutions for $x$ and $y$ into equation (\ref{eq:F4}) yields an expression for the wavefront aberration \cite{Born1}:
\begin{eqnarray}
    \fl\Psi(x_{o},x_{a},y_{o},y_{a})\nonumber\\
    =\frac{1}{p_{0}}\int_{z_{o}}^{z_{i}}F_{4}(x_{o},x_{a},y_{o},y_{a};z)\,dz,\nonumber\\
    =-\frac{1}{p_{0}}\int_{z_{o}}^{z_{i}}
        \biggl[
            \frac{A}{4}r_{o}^{4}
            +\frac{B}{4}r_{a}^{4}
            +C\kappa^{4}
            +\frac{D}{2}r_{o}^{2}r_{a}^{2}
            +Er_{o}^{2}\kappa^{2}
        \biggr.\nonumber\\
        \biggl.
            +Fr_{a}^{2}\kappa^{2}
        \biggr]dz,\label{eq:waveab}
\end{eqnarray}
where $z_{i}$ denotes the image plane. $r_{o}^{2}=x_{o}^{2}+y_{o}^{2}$, $r_{a}^{2}=x_{a}^{2}+y_{a}^{2}$ and $\kappa^{2}=x_{o}x_{a}+y_{o}y_{a}$ are rotational invariants.
The coefficients indicated by the capital letters indicate the following:
\begin{eqnarray}
    A=Ls^{4}+2Ms^{2}s'^{2}+Ns'^{4},\\
    B=Lt^{4}+2Mt^{2}t'^{2}+Nt'^{4},\label{eq:B}\\
    C=Ls^{2}t^{2}+2Mss'tt'+Ns'^{2}t'^{2},\\
    D=Ls^{2}t^{2}+M(s^{2}t'^{2}+s'^{2}t^{2})+Ns'^{2}t'^{2},\\
    E=Ls^{3}t+Mss'(st)'+Ns'^{3}t',\\
    F=Lst^{3}+Mtt'(st)'+Ns't'^{3}.
\end{eqnarray}
The aberration function $\Psi$ represents the deviation of the wavefront from the Gaussian wavefront at the aperture plane.
The five terms, except the $r_{o}^{4}$ term, which yields the offset of the entire phase, are known as the Seidel aberrations.
The deviation between the actual and Gaussian image points is denoted as the ray aberration.
This is expressed by taking the derivative of the wavefront aberration \cite{Born2}:
\begin{eqnarray}
    \Delta \mathbf{u}_{i}=
        (z_{i}-z_{a})\nabla_{a}\Psi(\mathbf{u}_{o},\mathbf{u}_{a}),
\end{eqnarray}
where $\mathbf{u}_{n}=(x_{n},y_{n})$ is a position vector at $z=z_{n}$ ($n=o,a,i$) and $\nabla_{a}=\partial/\partial x_{a}+\partial/\partial y_{a}$ represents a differential operator at the aperture plane.
Focusing on the $r_{a}^{4}$ term related to the SA among the five terms in equation (\ref{eq:waveab}), we obtain the ray aberration for the SA as follows:
\begin{eqnarray}
    \Delta \mathbf{u}_{i}=C^{(a)}_{S3}|\mathbf{u}_{a}|^{2}\mathbf{u}_{a},
\end{eqnarray}
where $C^{(a)}_{S3}$ denotes the third-order SA coefficient in the aperture representation, expressed as
\begin{eqnarray}
    C^{(a)}_{S3}=-\frac{(z_{i}-z_{a})}{p_{0}}\int_{z_{o}}^{z_{i}}B\,dz\label{eq:cs3a}.
\end{eqnarray}
To eliminate the variables related to the aperture plane from this expression, another pair of fundamental solutions, $g(z)$ and $h(z)$, are introduced.
These solutions obey the following boundary conditions at the object plane (figure \ref{fig:gh}):
\begin{eqnarray}
    \left\{
        \begin{array}{l}
            g(z_{o})=1\\
            g'(z_{o})=0
        \end{array}
    \right.,\quad
    \left\{
        \begin{array}{l}
            h(z_{o})=0\\
            h'(z_{o})=1
        \end{array}
    \right..\label{eq:gh}
\end{eqnarray}
Here, the plane in which $g(z)$ intersects the optical axis is the diffraction plane.
From geometric considerations, the following expressions can be obtained:
\begin{eqnarray}
    t(z)=t'_{o}h(z),\label{eq:tz}\\
    z_{i}-z_{a}=-1/t'_{i}=-\beta/t'_{o}\label{eq:ziza},
\end{eqnarray}
where $t'_{o}=t'(z_{o})$ and $t'_{i}=t'(z_{i})$.
$\beta=s(z_{i})=g(z_{i})$ denotes a lateral magnification of a lens system.
Now, the focus is only on the SA; thus, an object is assumed to be located on the electron optical axis.
Accordingly, the following expression is obtained:
\begin{eqnarray}
    \mathbf{u}_{a}=\mathbf{u}'_{o}/t'_{o}\label{eq:ua}.
\end{eqnarray}
Substituting equations (\ref{eq:B}), (\ref{eq:tz}), (\ref{eq:ziza}) and (\ref{eq:ua}) into (\ref{eq:cs3a}) yields:
\begin{eqnarray}
    \Delta \mathbf{u}_{i}=C'_{\mathrm{S3}}|\mathbf{u}'_{o}|^{2}\mathbf{u}'_{o},\\
    C'_{\mathrm{S3}}=\frac{\beta}{p_{0}}\int_{z_{o}}^{z_{i}}(Lh^{4}+2Mh^{2}h'^{2}+Nh'^{4})dz.
\end{eqnarray}
Note that the coefficient $C'_{\mathrm{S3}}$ of the round lens obtained using the electrostatic/magnetic field is always negative ($C'_{\mathrm{S3}}<0$) based Scherzer's theorem.
In electron optics, it is common to define the ray aberration at the object plane instead of the image plane; hence, the SA coefficient is expressed as $\Delta \mathbf{u}_{o}=\Delta \mathbf{u}_{i}/\beta$, where $\beta$ is generally negative.
The SA coefficient is obtained at the object plane as $C_{\mathrm{S3}}=C'_{\mathrm{S3}}/\beta>0$.
This is the reason the sign of the SA coefficient of the conventional electron round lens is considered to be always positive.

Assume that $|\mathbf{u}'_{o}|=r'_{o}\sim\alpha_{o}$ because the opening angle of the electron beam is generally small. Consequently, the expressions for the third-order SA are given by
\begin{eqnarray}
    \Delta r_{o}=C_{\mathrm{S3}}\alpha_{o}^{3},\\
    C_{\mathrm{S3}}=\frac{1}{p_{0}}\int_{z_{o}}^{z_{i}}(Lh^{4}+2Mh^{2}h'^{2}+Nh'^{4})dz\label{eq:cs3o}.
\end{eqnarray}
The aberration occurs in a refractive index medium; therefore, the integration interval in this equation can be replaced by the interval of the interaction length $L$ in which the lens action extends.
In the thin-lens approximation, the ray height does not change within the lens.
Hence, we can assume that $h=a$ and $h'= 0$.
Furthermore, the third-order SA coefficient is expressed as in equation (\ref{eq:cs3}).

The fifth-order SA coefficient is derived using the sixth-order expansion term.
Substituting the solutions (\ref{eq:xx}) and (\ref{eq:yy}) for $x$ and $y$ in equation (\ref{eq:F6}) to rewrite $F_{6}$ using the variables at the object and aperture planes and showing only the term for SA, the wave aberration function is obtained:
\begin{eqnarray}
    \fl\Psi_{\mathrm{S5}}(x_{o},x_{a},y_{o},y_{a})=\nonumber\\
    \frac{1}{p_{0}}\int_{z_{o}}^{z_{i}}\left(
        p_{6}t^{6}+\frac{p_{4}}{2}t^{4}t'^{2}-\frac{p_{2}}{8}t^{2}t'^{4}+\frac{p_{0}}{16}t'^{6}
    \right)r_{a}^{6}\,dz.
\end{eqnarray}
The same procedure used to derive equation (\ref{eq:waveab}) yields the following expression for the fifth-order SA coefficient:
\begin{eqnarray}
    \fl \Delta r_{o}=C_{\mathrm{S5}}\alpha_{o}^{5},\\
    \fl C_{\mathrm{S5}}=
        -\frac{1}{p_{0}}\int_{z_{o}}^{z_{i}}
    \biggl(
        6p_{6}h^{6}+3p_{4}h^{4}h'^{2}-\frac{3p_{2}}{4}h^{2}h'^{4}
    \biggr.\nonumber\\
    \biggl. 
        +\frac{3p_{0}}{8}h'^{6}
    \biggr)dz.
\end{eqnarray}
By further applying the thin-lens approximation to this expression, equation (\ref{eq:cs5}) can be achieved.

\end{document}